\documentstyle[twoside,fleqn,espcrc2]{article}

\catcode`@=11 
\def\@maketitle{
  \global\setbox\fm@box=\vbox\bgroup
   \preprintNumbers
    \vskip 8mm                    
    \raggedright                  
    \hyphenpenalty\@M             
    {\Large \@title \par}         
    \vskip\@bls                   
    {\normalsize                  
     \@author \par}               
    \vskip\@bls                   
    \@address                     
  \egroup
  \twocolumn[
    \unvbox\fm@box                
    \vskip\@bls                   
    \unvbox\abstract@box          
    \vskip 2pc]}                  

\def\address#1#2{\@makeadmark{#1}\begingroup
  \let\\=\relax
  \def\protect{\noexpand\protect\noexpand}\xdef\@address{\@address
  \protect\addresstext{#1}{#2}}\endgroup}
\catcode`@=12  

\newcommand{\AmS}{{\protect\the\textfont2
  A\kern-.1667em\lower.5ex\hbox{M}\kern-.125emS}}
\def\spa#1.#2{\left\langle\mskip-1mu#1\,#2\mskip-1mu\right\rangle}
\def\spb#1.#2{\left[\mskip-1mu#1\,#2\mskip-1mu\right]}

\title{Unitarity-based Techniques for One-Loop Calculations in QCD}

\author{Zvi Bern\address{a}{Dept. of Physics, University of California,
                         Los Angeles, CA 90095},
        Lance Dixon\address{b}{Stanford Linear Accelerator Center,
                            Stanford, CA 94309},
        and David A. Kosower\address{c}{Service de Physique Th\'eorique,
                                     Centre d'Etudes de Saclay,
                  F-91191 Gif-sur-Yvette cedex, France}\thanks{%
Presented at the 1996 Zeuthen Workshop on Elementary Particle Theory,
``QCD and QED in Higher Order'',
Rheinsberg, Brandenburg, April 21--26, 1996
}}

\begin{document}

\def\qb{{\overline q}}
\begin{abstract}
We discuss new techniques developed in recent years for performing
one-loop calculations in QCD, and present an example of results from the
process $0\rightarrow V q\qb g g$.
\end{abstract}

\def\preprintNumbers{%
\rightline{hep-ph/yyymmdd}
\rightline{UCLA/96/TEP/18}
\rightline{SLAC--PUB--7191}
\rightline{Saclay/SPhT--T96/071}}

\maketitle
\thispagestyle{empty}

\section{INTRODUCTION}

Perturbative QCD, and jet physics in particular, have matured sufficiently
that rather than being merely subjects of experimental studies, they are now
tools in the search for new physics.
This role can be seen in the search for the top quark,
as well as in recent speculations
about the implications of supposed high-$E_T$ deviations of the inclusive-jet
differential cross section at the Tevatron.
One of the important challenges to both theorists and experimenters
in coming years will be to hone jet physics as a tool in the quest
for physics underlying the standard model.
As such, it will be important
to measurements of parameters of the theory or of non-perturbative
quantities such as the parton distribution functions, as well as
to searches for new physics at the LHC.

Jet production, or jet production in association with identified photons
or electroweak vector bosons, appears likely to provide the best
information on the gluon distribution in the proton, and may also
provide useful information on the strong coupling $\alpha_s$.

In order to make use of these final states, we need a wider variety
of higher-order calculations of matrix elements.  Indeed, as we shall
review in the next section, next-to-leading order calculations are
in a certain sense the {\it minimal\/} useful ones.
On the other hand, these calculations are quite difficult with
conventional Feynman rules.
In following
sections, we will discuss some of the techniques developed in recent
years to simplify such calculations.

\section{NEXT-TO-LEADING ORDER CALCULATIONS}

While leading-order calculations often reproduce the shapes of
distributions well, they suffer from several practical and conceptual
problems whose resolution requires the use of next-to-leading order
calculations.  These problems are tied to the various logarithms
that can arise in perturbation theory.

The first of these logarithms are `UV' ones, connected with the
renormalization scale.  We are forced to introduce a renormalization scale
$\mu$ in order to define the coupling, $\alpha_s(\mu)$; but physical
quantities, such as cross sections or differential cross sections,
should be independent of $\mu$.  When we compute such a quantity
in perturbation theory, however, we necessarily truncate its expansion
in $\alpha_s$, and this introduces a spurious dependence on $\mu$.  This
dependence is significant in real-world applications of perturbative
QCD because the coupling is not that small, because it runs relatively
quickly, and because we are interested in processes with a relatively
large number of colored `final'-state partons.  Together, these effects
can lead to anywhere from a 30\% to a factor of 2--3 normalization uncertainty
in predictions of experimentally-measured distributions.
At leading order, the only dependence on $\mu$ comes from the resummation
of logarithms in the running coupling $\alpha_s(\mu)$, and the scale choice
is arbitrary.  At next-to-leading order, however, the virtual corrections
to the matrix element introduce another dependence on $\mu$.  This dependence
can --- and in practice, often does --- reduce the over-all sensitivity
of a prediction to variations in $\mu$.  (I should stress, however, that
while varying $\mu$ by some preset amount, say a factor of two up and
down from a typical scale, gives an indication of the sensitivity of or
uncertainty in the calculation, it does {\it not\/} give an estimate of
the error involved; for that one needs a next-to-next-to-leading order
calculation.)

The other logarithms are the `IR' ones, connected with the presence
of soft and collinear radiation.  Jets in a detector are not infinitely
narrow pencil beams; they consist of a spray of hadrons spread over
a finite segment.  Experimental measurements of jet distributions and
the like depend on resolution parameters, such as the jet cone size and
minimum transverse energy.  In a leading-order calculation, jets are
modelled by lone partons.  As a result, these predictions don't depend
on these parameters (or have an incorrect dependence on them).
In addition, the internal structure of a jet cannot be predicted at all.

At next-to-leading order, one necessarily includes contributions with
additional real radiation, which either shows up inside one of the jets,
or as soft radiation in the event.  This introduces, for
example, the required logarithmic dependence on the cone size $\Delta R$.

At a more conceptual level, we must remember that jet differential
cross sections are multi-scale quantities, involving not only a
hard scale (say a jet transverse energy $E_T$) but also scales
characterizing the resolution
defining a jet (for example, $E_T \Delta R$ or $E_{T{\rm min}}$).
As a result, the
perturbative expansion is not one in $\alpha_s$ alone, but contains
logarithms and logarithms-squared of ratios of scales.  These logarithms,
which arise from the infrared structure of gauge theories, might
spoil the applicability of perturbation theory if they grow too large.
Only a next-to-leading order calculation can tell us if we are safe.

\section{ORGANIZING THE CALCULATION OF AMPLITUDES}

A leading-order calculation of an $n$-jet process at a hadron collider
such as the Tevatron or LHC requires knowledge of the parton distribution
functions of the proton; of $\alpha_s$; and of the tree-level matrix
element for the $2\rightarrow n$ parton process.
A next-to-leading
order calculation of the same process requires in addition
the tree-level matrix element for the $2\rightarrow n+1$~parton process,
and the one-loop matrix element for the $2\rightarrow n$~parton process.
(In addition, it requires a general formalism such as that of
refs.~\cite{GG,GGK,Kunszt,Catani} for
cancelling the infrared singularities analytically while allowing a
numerical calculation of fully-differential observables.)

It is the calculation of the one-loop matrix element which presents
the greatest difficulty in the traditional Feynman-diagram approach.
This difficulty arises from the enormous number of diagrams, the large
amount of vertex algebra in each diagram, and the complexity of loop
integrals with many powers of the loop momentum in the numerator.  A
brute-force approach might easily lead to expressions thousands of times
larger than an appropriate representation of a result.

\def\oneloop{{\rm 1\hskip -1pt-\hskip -1ptloop}}
\def\tree{{\rm tree}}
\def\Gr{{\rm Gr}}
\def\Tr{{\rm Tr}}
\def\ibar{{\overline\imath}}
The first step in developing a more efficient method for performing
these calculations is to take advantage of the lessons from earlier
developments in tree-level calculations, and from string-based methods
for one-loop calculations.  These include
\begin{itemize}
\item Color decomposition: one should write the amplitude
as a sum over color factors multiplied by color-ordered subamplitudes,
and calculate these latter coefficients.  For a one-loop amplitude
for $0\rightarrow V q\qb g\cdots g$, the color decomposition takes the
form
\begin{eqnarray}
&&\hskip -15mm{\cal A}_{n+V}^\oneloop = g^n \sum_{j=1}^{n-1}
\sum_{\sigma\in S_{n-2}/ S_{n-2;j}} \Gr^{q\qb}_{n;j}(\sigma) \nonumber\\
   &&\hskip 0mm\times
    A^\oneloop_{n+V;j}(1_q,2_\qb;\sigma(3\!\cdots\! n); V)
\end{eqnarray}
where $\Gr^{q\qb}_{n;1}(3\cdots n)$ is the leading-color trace, equal
to the number of colors $N_c$ times the tree-level color factor
$(T^{a_3}\cdots T^{a_n})_{i_2}^{\ibar_1}$.  The remaining color factors
are subleading-color ones,
\begin{eqnarray}
&&\hskip -12mm\Gr^{q\qb}_{n;j>1}(3\cdots n) = \Tr(T^{a_3}\cdots T^{a_{j+1}})
  \nonumber\\
   &&\hskip 5mm\times
     (T^{a_{j+2}} \cdots T^{a_n})^{\ibar_1}_{i_2}\;.
\end{eqnarray}

\item
Spinor helicity method~\cite{SpinorHelicity}:
one should calculate helicity amplitudes
(and square or compute interference terms only numerically), using
spinor products $\spa{i}.j$ and $\spb{i}.j$.

\item
Use of supersymmetric decompositions and supersymmetry identities,
where applicable.

\item
Use of permutation identities~\cite{QQGGG} that allow rewriting
subleading-color
amplitudes as a sum of permutation of certain leading-color `primitive'
amplitudes.  One should then calculate only primitive amplitudes, which
correspond to leading-color amplitudes with a definite clockwise
or counter-clockwise orientation of internal fermion lines, and in which
the external fermion legs are not necessarily adjacent.
\end{itemize}

The reader will find a review of the techniques described below in
ref.~\cite{AnnReview}.

\section{PRIMITIVE AMPLITUDES}

In principle, one may imagine calculating primitive amplitudes
using color-ordered Feynman rules.  One can do better,
calculating them using
a combination of string-based rules and rules inspired by string theory.
Can one do better yet?  The unitarity-based technique, combined with judicious
use of the collinear limits, provides an even better way of calculating
these quantities.

In general, a one-loop amplitude has absorptive pieces; the basic
idea behind the unitarity-based technique
is that the cut-containing terms in the amplitude can be determined
by dispersion relations from the dispersive parts, which are in turn
related to a product of tree amplitudes.  As I will review below,
in supersymmetric theories, the basic technique in fact determines cut-free
pieces as well.

In calculating loop diagrams, we use dimensional regularization for both
the ultraviolet and infrared singularities; at one loop, it is possible
to choose a variant (dimension reduction)
 that preserves supersymmetry manifestly.  (The translation to the
conventional scheme is detailed in a paper by Kunszt, Signer, and
Tr\'ocs\'anyi~\cite{KST}.)

\section{INTEGRALS}

\def\e{\epsilon}
\def\IB{{\cal F}}
Let us consider the calculation of an $n$-point one-loop amplitude in
a gauge theory.  The calculation necessarily involves up to $n$-point one-loop
integrals; and given the powers of momentum in the three-gluon vertex,
may involve up to $m$ powers of the loop momentum in any $m$-point
integral.  Using either a combination of
Passarino-Veltman~\cite{PV} and van~Neerven-Vermaseren~\cite{vNV} reduction
techniques,
or equivalent ones~\cite{Integrals}, we can rewrite any such integral
(in $D=4-2\e$)
as a linear combination of scalar box integrals, scalar three-mass
triangles, scalar or tensor one- or two-mass triangles, and
scalar or tensor bubble integrals.  If we start with an amplitude
where all lines are massless, the internal lines of all these integrals
will also be massless; but the external legs will be sums of the
original external momenta, and thus possibly massive.  The scalar
boxes we must consider in general will thus have anywhere from
one to four external masses, for example.  We will denote the set
of integrals that appears in the computation of a $n$-point amplitude
by $\IB_n$.  The amplitude can be written as a sum over integrals
in this set, with coefficients $c_i$ which are rational functions
of the spinor products and Lorentz invariants.

\section{UNIQUE RECONSTRUCTION OF AMPLITUDES}

One of the basic tools in the method is the reconstruction theorem,
which tells us under which conditions the knowledge of the cuts
(in four dimensions)
suffices to reconstruct the amplitude uniquely.  It states that if,
for every loop integral
\begin{equation}
\int {d^D k\over (2\pi)^D}\,
  {{\rm Poly}(k^\mu)\over k^2 (k-p_1)^2 \cdots (k-p_{n-1})^2}
\end{equation}
which appears in the amplitude, the degree of the numerator polynomial
is less than or equal to $n-2$ (less than or equal to 1 if $n=2$), then
the amplitude is determined uniquely by its cuts in $D=4$.  This does
not imply that the amplitude is necessarily free of rational terms, but
merely that in this case, the rational terms are always linked to the
terms containing cuts.

Using the background-field method, and examining the effective action,
one can show that the theorem applies to the complete
amplitude in supersymmetric theories.  In $N=4$ supersymmetric
gauge theories,
one can go further, and show that in the background-field method, the
highest power of the loop momentum that can appear in the
numerator of an $n$-point integral is $n-4$; this in turn implies that
in these theories, the amplitude can be expressed as a sum of scalar box
integrals with rational coefficients $c_i$ (triangles and bubbles don't
enter).

\section{SEWING AMPLITUDES}

The key point in the unitarity-based method is that we sew
tree {\it amplitudes\/}, not tree diagrams.  This allows us to take
advantage of all the cancellations and reductions in numbers of
terms that have already occurred in
the process of computing the tree amplitudes.

A calculation using the unitarity-based method proceeds as follows.
We want to compute the coefficients $c_i$ of the integrals in
the set $\IB_n$.  To this end, we consider in turn cuts in all possible
channels.  For a given channel, we form the cut in that channel, summing
over all intermediate states; this gives rise to a phase-space
integral of the form
\begin{eqnarray}
&&\hskip -15mm\int d^D{\rm LIPS}(-\ell_1,\ell_2)\,
\nonumber\\
&&\hskip -5mm\times
A^\tree_L(-\ell_1,\ldots,\ell_2) A^\tree_R(-\ell_2,\ldots,\ell_1)\,,
\label{cutEqn}
\end{eqnarray}
where $\ell_{1,2}$ are the four-dimensional
on-shell momenta crossing the cut, and
where the $A^\tree_{L,R}$ are the color-ordered tree subamplitudes
on the two
sides of the cut.  Using the Cutkosky rules, we can rewrite this expression
as the absorptive or imaginary part of a loop amplitude,
\begin{equation}
\left[ \int {d^D\ell_1\over (2\pi)^D} A^\tree_L {1\over \ell_2^2+i\varepsilon}
     A^\tree_R {1\over \ell_1^2+i\varepsilon}\right]_{\rm cut}\,.
\end{equation}
Using the power-counting theorem in those cases where it applies
(or simply up to a rational ambiguity to be fixed later in cases where
it does not apply), we can recover the real parts by dropping the
subscript ``cut''.  We then perform the usual reductions on the
resulting loop integral, and extract the coefficient of any function
in $\IB_n$ containing a cut in the given channel.  (Functions not
containing a cut in the given channel should be dropped.)  Finally,
we reassemble the final answer by considering all channels.

Certain integrals, such as the box integrals, have cuts in more
than one channel.  Considering both channels provides us with a
cross check --- the coefficients as computed in both channels must
agree --- or alternatively we could reduce the amount of work we must
do by considering only one channel.  The latter choice is particularly
appropriate when computing amplitudes in an $N=4$ supersymmetric gauge theory,
since as noted above
all amplitudes in this theory can be written in terms of scalar
boxes.

\section{RATIONAL TERMS}

In nonsupersymmetric theories (such as QCD), there is a remaining
rational part which cannot be determined in this manner.  There are two
approaches to determining these pieces: the use of collinear limits,
and extending the cuts to ${\cal O}(\e)$.

\def\Split{{C}}
The collinear approach is based on the universal factorization of
amplitudes in the collinear limit,
\begin{eqnarray}
&&\hskip -10mm
A^\oneloop_{n;1}(\cdots,a,b) \mathop{\longrightarrow}^{a\parallel b}\nonumber\\
&& \hskip -5mm
 \sum_{\lambda_\Sigma=\pm} \Bigl[
  \Split^\tree_{-\lambda_\Sigma}(a^{\lambda_a},b^{\lambda_b})
      A^\oneloop_{n-1;1}(\cdots;\Sigma^{\lambda_\Sigma})\nonumber\\
&& \hskip 0mm
  +\Split^\oneloop_{-\lambda_\Sigma}(a^{\lambda_a},b^{\lambda_b})
      A^\tree_{n-1}(\cdots;\Sigma^{\lambda_\Sigma})\Bigr]
\end{eqnarray}
where the splitting amplitudes $\Split$ are universal functions depending
only on the collinear momenta and their helicities $\lambda$,
and on the momentum fraction $z$ ($k_\Sigma = k_a+k_b$, $k_a = z k_\Sigma$).
This limit is useful for deducing the rational pieces because the cut terms
alone will not produce the desired collinear limit; one must add rational terms
in order to obtain the correct limit.  However, there is no
proof that the terms deduced by this method are unique (though it is
likely true for more than five external legs; for five external legs,
there is an ambiguity arising from the existence of a term which contains
no cuts but is collinear-finite).  Such results thus need to be checked
(for example, numerically) against results from another method.

Another possible method again uses the cuts, but at ${\cal O}(\e)$ rather
than just to ${\cal O}(\e^0)$.  The basic point is that amplitudes in
a massless gauge theory have an over-all power of $(-s)^{-\e}$, where
$s$ is an invariant.  The ``rational'' pieces therefore do contain cuts
at ${\cal O}(\e)$, and can be deduced by sewing tree amplitudes, where
the momenta crossing the cuts are taken to be on-shell in $(4-2\e)$ dimensions
rather than four dimensions.  For scalars, this is equivalent to computing
massive rather than massless amplitudes, followed by an appropriately
weighted integration over
the ``mass'' (really the $(-2\e)$-dimensional component of the momentum).

\def\hf{{\textstyle{1\over2}}}
\def\Li{\mathop{\rm Li}\nolimits}
\def\qbar{{\overline q}}
\def\lr{\leftrightarrow}
\def\spab#1.#2.#3{\langle\mskip-1mu{#1}^-
                  | #2 | {#3}^-\mskip-1mu\rangle}
\def\spba#1.#2.#3{\langle\mskip-1mu{#1}^+
                  | #2 | {#3}^+\mskip-1mu\rangle}
\def\spaa#1.#2.#3{\langle\mskip-1mu{#1}^-
                  | #2 | {#3}^+\mskip-1mu\rangle}
\def\spbb#1.#2.#3{\langle\mskip-1mu{#1}^+
                  | #2 | {#3}^-\mskip-1mu\rangle}
\def\Th{\hat T}
\def\detprime{{\rm det}^\prime}
\def\rg{r_\Gamma}
\def\cg{c_\Gamma}
\def\li{{\rm Li_2}}
\def\Ls{\mathop{\rm Ls}\nolimits}
\def\Ll{\mathop{\rm L}\nolimits}
\def\Null{\mathop{}\nolimits}
\def\lv{\varepsilon}
\def\e{\epsilon}
\def\gluon{{\rm gluon}}
\def\gluino{{\tilde g}}
\def\qb{{\bar q}}
\def\susy{{\rm SUSY}}
\def\Ram{{\rm R}}
\def\Atree{A^{\rm tree}}
\def\tree{{\rm tree}}
\def\si{\sigma}
\def\ns{n_{\mskip-2mu s}}\def\nf{n_{\mskip-2mu f}}
\def\ib{{\bar\imath}}
\def\Split{\mathop{\rm Split}\nolimits}
\def\ul{\underline}
\def\gf{G_F}
\def\scalar{{\rm scalar}}

\section{AN ALL-MULTIPLICITY RESULT}

We had previously obtained\cite{Nfour}, with Dave Dunbar,
an explicit formula for the one-loop amplitude,
 in the $N=4$ supersymmetric gauge theory, with an
{\it arbitrary\/} number of external gluons with the helicity configuration
of the Parke-Taylor tree-level amplitudes~\cite{PT,MPreview},
namely all but two gluons
carrying the same helicity.  (Let us take the majority to have positive
helicity, the two opposite-helicity gluons to have negative helicity;
exchanging these two amounts to a complex conjugation.)
Here, we will outline the derivation of this result
using
 the unitarity-based method.
Such results demonstrate the power of the
method, because they would require the computation of an infinite
number of Feynman diagrams in the traditional approach.

This amplitude is uniquely determined by its cuts, and can be expressed
entirely in terms of scalar box integrals.  There are two kinds of cuts
to consider, one where both opposite-helicity gluons are one the same
side of the cut, the other where one such gluon appears on each side of
the cut.  The former cut receives contributions only from gluons crossing
the cut, the latter from fermions and scalars as well; but it turns out
that after summing over the $N=4$ supermultiplet (using supersymmetry
Ward identities~\cite{SusyWard}), the structure of the
second type of cut is similar to that of the first.  It thus suffices
to consider the first kind.

Let us suppose that the opposite-helicity gluons are on the left side
of the cut (drawn vertically through the diagram).  Then all the external
legs on the right-hand side of the cut have positive helicity; as a
result, the tree amplitude on the right-hand side will vanish unless
both legs crossing the cut on the right have negative helicity.  Since
we adopt the convention that all momenta are directed inwards into
an amplitude, the cut-crossing legs flip sign as we cross the cut to
the left.  We must flip their helicities as well, which tells us that
all legs of the tree amplitude to the left of the cut have
positive helicity, except the
two external legs carrying negative helicity.

Thus both tree amplitudes on either side of the cut have the helicity
configuration of the Parke-Taylor amplitudes~\cite{PT,MPreview},
so that simple all-$n$
formul\ae\ are available for them.  Most of the factors in these
amplitudes are independent of the momenta crossing the cut, and so
can be pulled out in front of the cut integral~(\ref{cutEqn}).  Indeed,
most of these factors reassemble into the tree amplitude itself, so
that (up to a phase) we obtain
\begin{eqnarray}
&& \hskip -15mm
A_n^\tree\;\int d^D{\rm LIPS}\;\nonumber\\
&&\hskip -10mm
 \times{\spa{(m_1-1)}.{m_1}\spa{\ell_1}.{\ell_2}^2 \spa{m_2}.{(m_2+1)}
    \over \spa{(m_1-1)}.{\ell_1}\spa{\ell_1}.{m_1}\spa{m_2}.{\ell_2}
          \spa{\ell_2}.{(m_2+1)}}
\end{eqnarray}
where $(m_1-1,m_1)$ and $(m_2,m_2+1)$ are the pairs of legs adjacent to
the cut.
The four factors in the denominator give rise to four propagators, which
along with the two propagators crossing the cut means that for an
arbitrary number of external gluons, we need consider only a hexagon
integral.  Indeed, it turns out to be a rather special hexagon integral;
rearranging the spinor products in its numerator, it decomposes into
a sum of one- and two-mass scalar boxes.  From this decomposition and
the uniqueness theorem, we then obtain the complete one-loop
$N=4$ supersymmetric amplitude.

\section{$Z$ DECAY TO FOUR JETS}

The decay of the $Z$ into jets has been studied extensively at LEP and SLC.
The three-jet to two-jet ratio is one of the better methods of
determining $\alpha_s(M_Z)$, but at present this extraction of the
running coupling is dominated by theoretical uncertainties.  The
resolution of these uncertainties requires a next-to-next-to-leading
order (NNLO) calculation of three-jet production.

Such a calculation will contain as ingredients the two-loop corrections
to $Z\rightarrow q\qb g$, which have yet to be computed;
the $Z\rightarrow q\qb ggg$ and $Z\rightarrow q\qb q'\qb' g$ matrix
elements at tree level, which are known~\cite{BGK}; and the
one-loop corrections to $Z\rightarrow q\qb gg$ and $Z\rightarrow q\qb q'\qb'$,
which we are currently computing.
(In addition, it will require an extension of one-loop techniques
for cancellation of IR singularities, such as the slicing method
of Giele and Glover~\cite{GG}, to NNLO.)

These one-loop corrections are also of interest outside of this NNLO
context, as they can be used to produce a program calculating
$Z\rightarrow4$~jets at next-to-leading order.  This process is the
lowest-order one in which the gluon and quark color charges can be
measured independently.  A next-to-leading order calculation will allow
reliable limits to be set, using LEP and SLC
data, on other possible light colored
fermions (such as light gluinos).

We present here the result for one of the leading-color subamplitudes
of the $Z\rightarrow q\qb gg$ process.
It is convenient to tack on the amplitude for the $Z$ production from two
leptons, so that all external legs are massless; in the formul\ae\ below,
these legs will be $k_5$ and $k_6$.  Furthermore, we will also include
a factor of $1/s_{56}$ (corresponding to a photon propagator replacing
the $Z$).

\def\args{(1^+_q,2^+,3^-,4^-_{\qbar};5^-_{\ell},6^+_{\ell})}
To express this result, we should first
record the corresponding tree amplitude,
\begin{eqnarray}
&&\hskip -10mm\Atree_{4+V}\args\ =\nonumber\\
&& \hskip -7mm i\,\Biggl[
  - { \spb1.2\spa1.3\spa4.5 \spba6.{(1+2)}.3
       \over \spa1.2 s_{23} t_{123} s_{56} }\nonumber\\
&&  \hskip -5mm + { \spa3.4\spb2.4\spb1.6 \spab5.{(3+4)}.2
       \over \spb3.4 s_{23} t_{234} s_{56} }\nonumber\\
&&   \hskip -5mm- { \spab5.{(3+4)}.2 \, \spba6.{(1+2)}.3
       \over \spa1.2 \spb3.4 s_{23} s_{56} } \Biggr]\,;\;
\end{eqnarray}
define several functions
(roughly corresponding to the various box and triangle integrals that
appear)
\begin{eqnarray}
&&\hskip -12mm
   \Ll_0(r) = {\ln(r)\over 1-r}\,,\hskip 5mm
  \Ll_1(r) = {\Ll_0(r)+1\over 1-r}\,,\hskip 5mm\nonumber\\
&&\hskip -12mm
  \Ls_{-1}(r_1,r_2) =
      \Li_2(1-r_1) + \Li_2(1-r_2)
   \nonumber\\
&&\hskip 0mm
 + \ln r_1\,\ln r_2 - {\pi^2\over6}\;,
   \nonumber\\
&&\hskip -12mm
  \Ls^{2{\rm m}h}_{-1}(s,t;m_1^2,m_2^2) =
    -\Li_2\left(1-{m_1^2\over t}\right)\\
&&\hskip 0mm
    -\Li_2\left(1-{m_2^2\over t}\right)
    -{1\over2}\ln^2\left({-s\over-t}\right)
      \nonumber\\
&&\hskip 0mm
    +{1\over2}\ln\left({-s\over-m_1^2}\right)
              \ln\left({-s\over-m_2^2}\right) \nonumber\\
&&\hskip 0mm
    + \biggl[ {1\over2} (s-m_1^2-m_2^2) + {m_1^2m_2^2\over t} \biggr]
      \nonumber\\
&&\hskip 5mm\times
        I_3^{3{\rm m}}(s,m_1^2,m_2^2) \;.\nonumber
\end{eqnarray}
as well as a flip symmetry,
\def\flip{{\rm flip}}
\begin{equation}
\hskip -5mm \flip = \{1 \lr 4,\, 2 \lr 3,\, 5 \lr 6;
\, \spa{\,}.{\,} \lr \spb{\,}.{\,}\}\,.
\end{equation}
The one-loop amplitude is then
\begin{equation}
A^\oneloop_{4+V;1} = V_a \Atree_{4+V} + F_g + F_s
\end{equation}
where
\begin{eqnarray}
&&\hskip -13mm
V_a =
 - {1\over\e^2} \Biggl[
      \left({\mu^2\over-s_{12}}\right)^\e
    \!+\! \left({\mu^2\over-s_{23}}\right)^\e
    \!+\! \left({\mu^2\over-s_{34}}\right)^\e \Biggr]\nonumber\\
&&
 - {3\over2\e} \left({\mu^2\over-s_{56}}\right)^\e - {7\over2}\,.
\end{eqnarray}
and where
\small
\begin{eqnarray}
&&\hskip -17mm F_g =
  B_1 \Ls_{-1}\Bigl({-s_{23}\over-t_{234}},{-s_{34}\over-t_{234}}\Bigr)
+ B_5 \Ls_{-1}\Bigl({-s_{12}\over-t_{123}},{-s_{23}\over-t_{123}}\Bigr)
  \nonumber\\
&&\hskip -14mm
+ B_2 \Ls^{2{\rm m}h}_{-1}(s_{34},t_{123};s_{56},s_{12})
 + T \, I_3^{3{\rm m}}(s_{12},s_{34},s_{56})
\nonumber\\
&&\hskip -14mm
+ B_4 \Ls^{2{\rm m}h}_{-1}(s_{12},t_{234};s_{56},s_{34})
\nonumber\\
&&\hskip -14mm
- 2\,{ \spa1.3\spba6.{(1+2)}.3 \over \spa1.2\spb5.6\,\spba4.{(2+3)}.1 }
  \\
&&\hskip -15mm
  \times\Biggl[ \!
   { \spbb6.{(2\!+\!3)1}.2 \over t_{123} }\!
    { \Ll_0\Bigl(\!{-s_{23}\over-t_{123}}\!\Bigr) \over t_{123} }
 \!+\! { \spb6.4\!\spa4.3 \over \spa2.3 }\!
    { \Ll_0\Bigl(\!{-s_{56}\over-t_{123}}\!\Bigr) \over t_{123} }
        \Biggr] \nonumber\\
&&\hskip -14mm
- 2\,{ \spb4.2\spab5.{(3+4)}.2 \over \spb4.3\spa6.5\,\spab1.{(2+3)}.4 }
 \nonumber\\
&&\hskip -15mm
  \times\Biggl[\!
   { \spaa5.{(2\!+\!3)4}.3 \over t_{234} }\!
    { \Ll_0\Bigl(\!{-s_{23}\over-t_{234}}\!\Bigr) \over t_{234} }
 + { \spa5.1\!\spb1.2 \over \spb3.2 }\!
    { \Ll_0\Bigl(\!{-s_{56}\over-t_{234}}\!\Bigr) \over t_{234} }
        \Biggr]\, , \nonumber
\end{eqnarray}
and
\begin{eqnarray}
&&\hskip -14mm F_{s1} =
 - {1\over2} { \spa1.3
     \over \spa1.2\spa2.3\spb5.6 t_{123} \, \spba4.{(2+3)}.1 }
  \nonumber\\
&&\hskip -10mm \times
  \Biggl[
  \bigl( \spba6.{(2+3)12}.3 \bigr)^2
    { \Ll_1\Bigl({-t_{123}\over-s_{23}}\Bigr) \over s_{23}^2 }
  \nonumber\\
&&\hskip -7mm
  + \bigl( \spb6.4\spa4.3 t_{123} \bigr)^2
    { \Ll_1\Bigl({-s_{56}\over-t_{123}}\Bigr) \over t_{123}^2 }
                \Biggr] \\
&&\hskip -10mm
+ {1\over2} { {\spb6.2}^2
    \over \spa1.2\spb2.3\spb3.4\spb5.6 } \nonumber\\
&&\hskip -14mm
F_s = F_{s1} + \flip(F_{s1})\,.\nonumber
\end{eqnarray}

In these expressions,
\begin{eqnarray}
&&\hskip -12mm
 B_5 =
  { \spa1.3 \bigl( \spba6.{(1+2)}.3 \bigr)^2
  \over \spa1.2\spa2.3\spb5.6 t_{123} \, \spba4.{(2+3)}.1 }  \nonumber\\
&&\hskip -8mm
+ { {\spb1.2}^3 {\spa4.5}^2
  \over \spb2.3\spb1.3\spa5.6 t_{123} \, \spba1.{(2+3)}.4 }\ .\nonumber\\
&&\hskip -12mm
 B_2 =
  { \spa1.3 \bigl( \spba6.{(1+2)}.3 \bigr)^2
    \over \spa1.2\spa2.3\spb5.6 t_{123} \, \spba4.{(2+3)}.1 }  \\
&&\hskip -8mm
+ { {\spb1.2}^2 {\spa4.5}^2 \, \spba2.{(1+3)}.4
    \over \spb2.3\spa5.6 t_{123}\,\spba1.{(2+3)}.4\,\spba3.{(1+2)}.4 }
    \ , \nonumber\\
&&\hskip -12mm
 B_1 = \flip(B_5)\ , \nonumber\\
&&\hskip -12mm
 B_4 = \flip(B_2)\ , \nonumber
\end{eqnarray}
and
\begin{eqnarray}
&&\hskip -12mm T_1 =
  { t_{123} \, (s_{56}+s_{12}-s_{34}) - 2s_{12}s_{56}
    \over 2 t_{123} } \, B_2  \\
&&\hskip -7mm
+ {1\over2} { \spb1.2 \Bigl[ \bigl( \spaa4.{(1+2)(3+4)}.5 \bigr)^2
                             - s_{12} s_{34} \spa4.5^2 \Bigr]
              \over \spa1.2\spb3.4\spa5.6
                    \spba1.{(2+3)}.4 \spba3.{(1+2)}.4 } \nonumber\\
&&\hskip -12mm
 T = T_1 + \flip(T_1)\;.\nonumber
\end{eqnarray}

\normalsize
  This amplitude is one of three leading-color helicity amplitudes; in
addition, five other primitive helicity amplitudes are needed to construct
the subleading-color amplitudes for $Z\rightarrow$~four jets.

The work described above was
supported in part by the US Department of Energy under grants
DE-FG03-91ER40662 and DE-AC03-76SF00515, by the Alfred P. Sloan
Foundation under grant BR-3222, and
by a NATO Collaborative Research Grant CRG--921322 (L.D. and D.A.K.).

\end{document}